\documentclass[conference]{IEEEtran}
\IEEEoverridecommandlockouts
\usepackage{cite}
\usepackage{amsmath,amssymb,amsfonts}
\usepackage{algorithmic}
\usepackage{array,multirow,graphicx}
\usepackage{textcomp}
\usepackage{xcolor}
\usepackage{soul}
\usepackage{rotating}
\usepackage{pifont}
\def\BibTeX{{\rm B\kern-.05em{\sc i\kern-.025em b}\kern-.08em
    T\kern-.1667em\lower.7ex\hbox{E}\kern-.125emX}}
\begin{document}

\title{Analyzing Machine Learning Approaches for Online Malware Detection in Cloud}

\author{\IEEEauthorblockN{Jeffrey C Kimmell\IEEEauthorrefmark{1}, Mahmoud Abdelsalam\IEEEauthorrefmark{2}, and Maanak Gupta\IEEEauthorrefmark{3}}
\IEEEauthorblockA{\IEEEauthorrefmark{1}\IEEEauthorrefmark{3}{Dept. of Computer Science},
{Tennessee Technological University},
Cookeville, Tennessee 38505, USA \\\IEEEauthorrefmark{2}Dept. of Computer Science, Manhattan College,
Riverdale, NY, USA\\}
\IEEEauthorrefmark{1}jckimmell42@tntech.edu, 
\IEEEauthorrefmark{2}mabdelsalam01@manhattan.edu,
\IEEEauthorrefmark{3}mgupta@tntech.edu}

\maketitle

\begin{abstract}
The variety of services and functionality offered by various cloud service providers (CSP) have exploded lately.
Utilizing such services has created numerous opportunities for enterprises infrastructure to become cloud-based and, in turn, assisted the enterprises to easily and flexibly offer services to their customers. The practice of renting out access to servers to clients for computing and storage purposes is known as Infrastructure as a Service (IaaS).
The popularity of IaaS has led to serious and critical concerns with respect to the cyber security and privacy. In particular, malware is often leveraged by malicious entities against cloud services to compromise sensitive data or to obstruct their functionality. In response to this growing menace, malware detection for cloud environments has become a widely researched topic with numerous methods being proposed and deployed. In this paper, we present online malware detection based on process level performance metrics, and analyze the effectiveness of different baseline machine learning models including, Support Vector Classifier (SVC), Random Forest Classifier (RFC), K-Nearest Neighbor (KNN), Gradient Boosted Classifier (GBC), Gaussian Naive Bayes (GNB) and Convolutional Neural Networks (CNN). Our analysis conclude that neural network models can most accurately detect the impact malware have on the process level features of virtual machines in the cloud, and therefore are best suited to detect them. Our models were trained, validated, and tested by using a dataset of 40,680 malicious and benign samples.
The dataset was complied by running different families of malware (collected from VirusTotal) in a live cloud environment and collecting the process level features.

\end{abstract}

\section{Introduction and Motivation}
Cloud computing's convenience and scalability have made it the go to resource for many entities in both private and public sectors.
One of the major cloud characteristics is offering resources on-demand, if and when needed using the pay-as you go model. In general, IT specialists at enterprises have to control and manage these resources which hinders the advantages offered by the cloud. Cloud automation tools have become the norm where IT personnel are able to automatically provision resources in the cloud. Such automation is achieved through tools (e.g., Puppet\footnote{Puppet. https://puppet.com/} and Chef\footnote{Chef. https://www.chef.io/}) by writing configuration scripts that are able to create, modify, and delete resources in the cloud.
Just as such orchestration tools offer huge benefits to DevOps teams, they widen the security attack surface. 
In particular, VMs are often spawned using automatic configuration tools which means that a large group VM are similarly configured, if not exact copies. The inherent redundancies in these VMs could allow for malware to easily propagate across VMs, especially if there are flaws in these configuration scripts. The repercussions of compromising a group of VMs far outweighs those of a single compromised VM.
Cloud infrastructure requires considerably major security implementations due to its inherent complexity and dynamic environment where threats are always changing and evolving. For the same reason, developing malware detection methods that are both accurate and fast is imperative \cite{watson2015malware}.  

Malware is a major threat to cloud infrastructures. Multiple malware detection methods have been proposed with pros and cons.
Static malware detection \cite{alazab2010zero, shalaginov2018machine, shalaginov2018machine, roseline2020intelligent} is a popular method where the signature of an executable is analyzed and  compared to a database of known malware signatures. Attackers have tried to limit the effectiveness of static analysis by implementing techniques such as obfuscation and packing. In addition, static malware analysis is limited to known malware executables and is unable to detect the ever-evolving zero-day malware. These two major limitations have led to extensive research on behavioral malware detection methods. 
Dynamic and online malware detection are two behavioral based methods. Dynamic malware detection methods work by running the malware executables in a secure environment, such as a sandbox and analyzing their behavior. By doing this, the detection system is able to analyze novel zero-day malware since it is not relying on previously known signatures but the actual behavior of the executable. 
However, attackers have been able to implement malware that can detect the use of tools such as a sandbox, and cease behaving maliciously in order to avoid its detection. Dynamic and static methods also share the same limitation where the detection system focus on identifying malware in the given executables before they are run on actual systems.
However, it is common for malware to get into a system through vulnerabilities, hence bypassing these primitive detection approaches. Online malware detection \cite{abdelsalam2019onlinemalware,abdelsalam2017clustering,abdelsalam2018malwaremalwarecnn,mcdole2020analyzing} focuses on the behavior of a machine that it is trying to protect from malware. Rather than analyzing executables and their behavior, online methods monitor the performance of the entire virtual machine, and raises an alert if traces of malicious behavior is found at any time. As such, online malware detection methods are considered continuous real-time detection system, and overcomes the limitations of static and dynamic malware detection approaches.

Machine Learning (ML) and neural networks techniques are widely utilized in order to capture the behavior of malware in an accurate and efficient way~\cite{tobiyama2016malware}. This is due to the models' abilities to quickly process a significant amount of data generated by a VM to classify executables as malicious or benign.  
Online malware detection techniques are significantly impacted by the features chosen to capture the behavior of the malware present in a particular machine. For example, several works~\cite{alazab2010zero, dawson2018phase, luckett2016neural} are using system calls (the most widely used); however, it is very resource consuming and data can only be fetched through on-host collecting agent. Only few works~\cite{abdelsalam2019onlinemalware,abdelsalam2017clustering,abdelsalam2018malwaremalwarecnn, watson2015malware,mcdole2020analyzing,mcdole2020deep,9422693} use resource utilization metrics (also known as performance metrics) due to the fact that they are less expressive than system calls in terms of capturing the low-activity malicious behavior. However, performance metrics are more suitable to cloud environments since they are \textit{cheaper} and can easily be fetched from the hypervisor (e.g., VMs introspection~\cite{garfinkel2003virtual}).

In this paper, we analyze and compare the effectiveness of different online malware detection approaches that utilize process-level performance metrics. We provide an in depth analysis of various machine learning models which will work as a baseline for other works which focus primarily on one machine learning model. This is critical to motivate the use of \textit{expensive} deep learning models which require huge amounts of training data and to prove their efficacy with respect to more fundamental machine learning models. To our understanding and literature review, this is the first work focusing on analyzing the efficiency of baseline machine learning models, which is important to \textit{justify} the use of expensive deep learning techniques.


The \textit{main contributions} of this paper are as follows:
\begin{itemize}[]
    \item We analyze the effectiveness of different machine learning models for online malware detection.
    \item We demonstrate how the set of processes running in a VM can be represented as a sequence of system features.
    \item We conjecture that a Convolutional Neural Network (CNN) model is better suited for malware detection compared to traditional ML methods.
\end{itemize}

The remainder of the paper is structured as follows. Section \ref{sec:related} discusses the related works regarding cloud malware detection and use of various machine learning models. We also elaborate different ML models used in our work. Section \ref{sec:setup} discusses the experimental cloud set up and methodology. The results generated by each of the models are discussed in Section \ref{subsec:Results}. Section \ref{sec:analysis} offers comparison and analysis of different ML approaches used followed by limitations in Section \ref{sec:limitation}. Section \ref{sec:conclusion} summarizes our work.

\section{Background}
\label{sec:related}
We will outline related work in malware detection and summarize different ML models used in our work.

\subsection{Related Works}

\begin{table*}[!t]
    \caption{This table shows the differences between our work and related works. The sections in \textcolor{red}{red} indicate a difference in features used, environment that the solution was tested in, or focus of the paper. A \ding{51} indicates that a particular paper possesses this attribute or model, whereas a blank cell indicates an absence of this attribute or model.}
    \centering
    \begin{tabular}{| c | c | c | c | c | c || c | c || c | c | c || c | c | c | c | c | c | c | c | c |}
    \hline
    & \multicolumn{5}{ |c|| }{Features}
    & \multicolumn{2}{ |c|| }{Domain}
    & \multicolumn{3}{ |c|| }{Focus}
    & \multicolumn{9}{ |c| }{Models Used} \\
    \hline
    {Paper} & 
    {\rotatebox[origin=c]{90}{\textcolor{red}{API Calls}}} & 
    {\rotatebox[origin=c]{90}{\textcolor{red}{Performance Metrics}}}&
    {\rotatebox[origin=c]{90}{\textcolor{red}{System Calls}}}&
    {\rotatebox[origin=c]{90}{\textcolor{red}{Performance Counters}}}&
    {\rotatebox[origin=c]{90}{\textcolor{red}{Memory Features}}}&
    {\rotatebox[origin=c]{90}{\textcolor{red}{Cloud Environment}}} &
    {\rotatebox[origin=c]{90}{\textcolor{red}{Traditional Host-Based Environment}}} &  
    {\rotatebox[origin=c]{90}{\textcolor{red}{Dynamic Malware Detection}}}&
    {\rotatebox[origin=c]{90}{\textcolor{red}{Online Malware Detection}}}&
    {\rotatebox[origin=c]{90}{\textcolor{red}{Anomaly Detection}}}&
    {\rotatebox[origin=c]{90}{KNN}} & 
    {\rotatebox[origin=c]{90}{Naive Bayes}} & 
    {\rotatebox[origin=c]{90}{Neural Network}} & 
    {\rotatebox[origin=c]{90}{Random Forest}} &
    {\rotatebox[origin=c]{90}{Boosted Trees}} & 
    {\rotatebox[origin=c]{90}{SVC}} & 
    {\rotatebox[origin=c]{90}{Clustering}} & 
    {\rotatebox[origin=c]{90}{Decision Trees}} &
    {\rotatebox[origin=c]{90}{No Machine Learning}}\\ [.5ex] 
    \hline
    
    
    Firdausi et al. 2010 \cite{firdausi2010analysismalwareml} & 
     & 
     &
    \ding{51} &
     &
     &
     & 
    \ding{51} &
    \ding{51}& 
    &
    &
    \ding{51} & 
    \ding{51} & 
     &
     &
     &
    \ding{51} &
     &
    \ding{51} &
    \\ 
    \hline
    
    Azmandian et al. 2011 \cite{azmandian2011virtual} & 
     & 
     &
    \ding{51} &
     &
     &
    \ding{51} & 
     &
     & 
    \ding{51}&
    \ding{51}& 
    \ding{51} & 
     & 
     & 
     & 
     &
     &
    \ding{51} & 
     &
    \\ 
    \hline
    
    Guan et al. 2012 \cite{guan2012ensemble} & 
     & 
    \ding{51} &
     &
     &
     &
    \ding{51} & 
     &
    & 
    \ding{51} &
    \ding{51} &
     &
    \ding{51} &
     &
     & 
     & 
     &
     &
    \ding{51} &
    \\ 
    \hline
    
    Pannu et al. 2012 \cite{pannu2012aad} & 
     & 
    \ding{51} &
     &
     &
     &
    \ding{51} & 
     &
    & 
    &
    \ding{51} &
     &
     &
     &
     & 
     & 
    \ding{51} &
     &
     &
    \\ 
    \hline
    
    Demme et al. 2013 \cite{demme2013feasibility} &
     &
     &
     &
    \ding{51} &
     &
     &
    \ding{51} &
    & 
    \ding{51}&
    &
    \ding{51} & 
     & 
     & 
    \ding{51} & 
     & 
     & 
     & 
    \ding{51} &
    \\ 
    \hline
    
    Pirscoveanu et al. 2015 \cite{pirscoveanu2015analysis} & 
     & 
    \ding{51} &
     &
     &
     &
     & 
    \ding{51} &
    \ding{51}& 
    & 
    &
     & 
     & 
     & 
    \ding{51} & 
     & 
     & 
     & 
     &
    \\ 
    \hline
    
    Watson et al. 2015 \cite{watson2015malware}  & 
     & 
    \ding{51}  & 
     &
     &
     &
    \ding{51} & 
     &
    & 
    \ding{51} &
    \ding{51} &
     & 
     & 
     &  
     &  
     & 
    \ding{51} & 
     & 
     &
    \\ 
    \hline
    
    Luckett et al. 2016 \cite{luckett2016neural} & 
     &
     &
    \ding{51} &
     &
     &
     & 
    \ding{51} &
    \ding{51} & 
    &
    &
     &
     & 
    \ding{51} &
     & 
     & 
     & 
     & 
     &
    \\ 
    \hline
    
    Fan et al. 2016 \cite{fan2016maliciousmalware} &
    \ding{51} &
     &
     &
     &
     &
     &
    \ding{51} &
    \ding{51}& 
    &
    &
    \ding{51} &
     &
     & 
     & 
     & 
     &
     & 
     &
    \\ 
    \hline
    
    Tobiyama et al. 2016 \cite{tobiyama2016malware} & 
    \ding{51} &
     &
     &
     &
     &
     &
    \ding{51} &
    \ding{51} & 
    &
    &
     & 
     & 
    \ding{51} & 
     & 
     & 
     & 
     &
     &
    \\ 
    \hline
    
    Abdelsalam et al. 2017\cite{abdelsalam2017clustering} & 
     &
    \ding{51} &
     &
     &
     &
    \ding{51} &
     &
    & 
    \ding{51} &
    \ding{51}&
     &
     &
     & 
     & 
     &
     &
    \ding{51} & 
     &
    \\ 
    \hline
    
    Xu et al. 2017 \cite{xu2017malware} &
     & 
     &
     &
     &
    \ding{51} &
     & 
    \ding{51} &
    & 
    \ding{51} &
    &
     &
     & 
     & 
    \ding{51} & 
     & 
     & 
     & 
     &
    \\ 
    \hline
    
    Abdelsalam et al. 2018 \cite{abdelsalam2018malwaremalwarecnn} &
     &
    \ding{51} &
     &
     &
     &
    \ding{51} & 
     &
    & 
    \ding{51} &
    &
     & 
     & 
    \ding{51} & 
     & 
     & 
     & 
     & 
     &
    \\ 
    \hline

    Dawson et al. 2018 \cite{dawson2018phase} &
     &
     &
    \ding{51} &
     &
     &
    \ding{51} & 
     &
    & 
    \ding{51} &
    \ding{51} &
     &
     &
    \ding{51} & 
     & 
     & 
     & 
     &
     &
    \ding{51}\\ 
    \hline
    
    Joshi et al. 2018 \cite{joshi2018machine} &
     & 
    \ding{51} &
     &
     &
     &
     &
    \ding{51} &
    \ding{51}& 
    &
    &
     & 
     & 
     &
    \ding{51} &
     & 
     &
     & 
     &
    \\ 
    \hline
    
    
    
    \textbf{Our Approach} &
    &
    \ding{51} &
    &
    &
    &
    \ding{51}&
    &
    & 
    \ding{51} &
    &
    \ding{51} &
    \ding{51} & 
    \ding{51} &
    \ding{51} &
    \ding{51} & 
    \ding{51} &
    &
    &\\ 
    \hline
    \end{tabular}
    \label{table:related_works}
\end{table*}

There has been substantial work in the field of malware detection. Most recently, approaches that rely on machine learning techniques have gained traction. The high increase in cloud activity has also called for more attention towards methods that are specific to the cloud environment \cite{abdelsalam2018malwaremalwarecnn, abdelsalam2017clustering, abdelsalam2019onlinemalware, watson2015malware, dawson2018phase, pirscoveanu2015analysis, fan2016maliciousmalware}. 
Table~\ref{table:related_works} shows some of the closely related work. We categorize the work with respect to the focus of the paper, the targeted environment, and the features used for detection as well as the ML algorithms used.


\textbf{Dynamic Malware Detection.} Dynamic malware detection approaches focus on running malware executables in a sandbox and closely monitor its behavior or system wide behavior. Most works target traditional host-based systems. Research in \cite{firdausi2010analysismalwareml, luckett2016neural} utilize system calls as features to train classical machine learning models (i.e. KNN, NB, SVC and DT) and neural networks, respectively. Other works \cite{tobiyama2016malware, fan2016maliciousmalware, pirscoveanu2015analysis} analyze the effectiveness of using CNN, RF and KNN models for malware detection and rely on features extracted from API calls. In addition, Joshi et al. \cite{joshi2018machine} use the random forest classifier and monitor a VM's process behavior. However, their analysis of these approaches is not analyzed on a cloud environment.
The main limitation of such approaches is the fact that they are performed in an isolated environment neglecting the unique cloud topology, including the infrastructure and its network communication channels. Even though dynamic analysis approaches can be adopted and used in online settings, collecting real-time metrics generated from the cloud environment is essential for cloud malware detection.
 
 \textbf{Online Malware Detection.} Unlike static and dynamic analysis approaches where an executable is analyzed or monitored before it runs on a system, online malware detection approaches focus on continuously monitoring the entire systems under the assumption that a malware will eventually make its way into the system. The authors in \cite{demme2013feasibility} introduced a malware detection method that utilizes performance counters and \cite{xu2017malware} proposed the use of memory features; however, both of these works are targeting traditional host-based environment. Other works specifically target the cloud. Abdelsalam et al. \cite{abdelsalam2018malwaremalwarecnn} presented a CNN solution that focused on process level performance metrics with a relatively successful accuracy score of 90\%. However, this work only examines CNN and does not provide a baseline of comparison with respect to traditional machine learning algorithm, which we aim to accomplish in this paper. In addition, we also categorize anomaly detection based approaches as online techniques, since they naturally focus on continuous monitoring of their target systems. Pannu et al. \cite{pannu2012aad} use cloud performance metrics features and analyzed the effectiveness SVM and Gaussian based approaches. Even though their work focused on general anomaly detection within the cloud, it can be easily adopted and tailored to detect malware specifically. Similarly, Guan et al. \cite{guan2012ensemble} considered anomaly detection within a cloud environment where they analyzed system calls based on an ensemble of Bayesian predictors and decision trees. Their work also focused on cloud computing systems failure, not specifically malware. Other anomaly detection based approaches are focused on malware. Azmandian et al. \cite{azmandian2011virtual} presented an intrusion detection system using system calls as features. Abdelsalam et al. \cite{abdelsalam2017clustering} proposed a novel k-means clustering algorithm for detection purposes. This approach succeeded in detecting highly active malware, but was not successful in detecting low activity malware. Dawson et al. \cite{dawson2018phase} fetch API calls through hypervisor to be used as features and use a non linear phase-space algorithm to detect anomalous behavior. Watson et al. \cite{watson2015malware} use performance metrics to build a one class SVC; however, the authors experimented on highly active malware which is easy to detect.

In this paper, we aim to address the following limitations:
\begin{enumerate}
    \item Unlike traditional host-based approaches in \cite{firdausi2010analysismalwareml, demme2013feasibility, pirscoveanu2015analysis, luckett2016neural, fan2016maliciousmalware, tobiyama2016malware, xu2017malware, joshi2018machine}, we aim to focus on developing a cloud-specific approach. Our experiment deployment, which consists of a commonly used three-tier web architecture, gives our collected data the added benefit of being generated in an extremely realistic cloud environment. The different layers of this architecture allows the utilization of the cloud topology and provides an in-depth view at how a real system could be affected by malware.
    \item Unlike dynamic analysis approaches in \cite{firdausi2010analysismalwareml, pirscoveanu2015analysis, luckett2016neural, fan2016maliciousmalware, tobiyama2016malware}, we aim to focus on developing an online malware detection approach that is well suited for cloud environments. 
    \item Unlike the work in \cite{firdausi2010analysismalwareml, azmandian2011virtual, demme2013feasibility, luckett2016neural, fan2016maliciousmalware, tobiyama2016malware, xu2017malware, dawson2018phase}, we focus on performance-level metrics which is more practical for cloud than costly features like system or API calls.
    \item Unlike the work in \cite{watson2015malware}, we aim to focus on a broad range of malware by using low-active malware from seven different categories.
    \item Unlike the work in \cite{abdelsalam2018malwaremalwarecnn}, we aim to provide a baseline comparison by employing various traditional machine learning algorithms to convey the importance of using deep learning algorithms for online malware detection in cloud.
\end{enumerate}



\subsection{Baseline Machine Learning Models}
\label{sec:method}
Here we outline different ML models used in our analysis.
\subsubsection{Support Vector Classifier (SVC)}
SVC are supervised learning models that are used for classification. SVC's ability to use a non-linear kernel gives this method the ability to efficiently perform non-linear classifications. This also reduces the computational power required to calculate relationships in infinite dimensions. There is not always a possible linear classification discernible between features, so finding higher dimensional relationships between the supplied features allows SVC to make classifications that other methods, such as logistic regression, would not be able to make.

\subsubsection{Random Forest Classifier (RFC)}
RFC are supervised learning models used for classification. An RFC works by fitting a collection of decision trees\cite{breiman1984classification}, and is therefore considered an ensemble learning method since it uses a collection of classifiers \cite{breiman2001random}. RFCs choose the best parameter at each node at random as opposed to decision trees where the best parameter is selected based on all of the features \cite{joshi2018machine}. This gives RFCs better scalability as well as reducing the risk of overfitting.

\subsubsection{Nearest Neighbor}

Also known as k-Nearest Neighbor (KNN), it is a supervised learning method of classification that relies on measuring the distances of samples that have close proximity. KNN assumes that samples with the same classification will be closer in distance, and uses this assumption to  classify a new sample based on the k closest neighbors.

\subsubsection{Gradient Boosted Trees}
Gradient Boosted Trees, or Gradient Boosted Classifiers (GBC), like RFC, is an ensemble supervised learning method. GBCs work by creating many decision trees that handle specific decisions. A weak learner is considered a model that is only slightly better than guessing. However, weak learners are often able to make correct decisions on a very specific portion of the sample. The idea is to create an ensemble of enough weak learners so that the model as a whole can uses the decisions from the various learners to generate an overall accurate classification. 

\subsubsection{Naive Bayes}
A Naive Bayes classifier generates classifications using the Bayes Theorem. Bayes Theorem is a method used to calculate conditional probability based off a set of features, however it requires a large pool of computational resources. Bayes Theorem assumes that all features are dependent on one another, this is what leads to the theorem becoming so computationally demanding. To remedy this, a simplified or \textit{Naive} method was created by assuming that each of the features are independent; this assumption allows the theorem to be simplified which therefore reduces the required computational resources.

\subsubsection{CNN}
Convolutional Neural Networks are most commonly used for image recognition or use cases that involve visual data. We can imitate this with our data by shaping it so that it is represented in a two dimensional array, similar to an image. The CNN model that we employ in our paper is DenseNet-121 model, which is one of the state-of-the-art CNN models. 

DenseNets \cite{huang2017densely} attempt to solve the issue of the "vanishing gradient." This problem arises when the neural network becomes so deep that the standard back-propagation fails to update the neurons at the early layers of the network from the changes made at the output. Because of this issue, neural networks were limited in deepness and complexity. DenseNets work around this issue by creating more channels that connect the hidden layers. In DenseNets, the outputs of every layer are passed to subsequent layers. Because of this, DenseNets do not need as many feature maps at every hidden layer since these feature maps are being used by every subsequent layer and not just the successor layer. In addition, DenseNets borrow the identity mapping feature from residual networks. This feature allows the gradient to flow through the model easily via the use of skip-connections. DenseNets are comprised of dense blocks and connected with transition layers that are made of a convolution and pooling layer.
For simplicity, we will refer to it as CNN for the remainder of the paper.

\section{Methodology and Experimental Setup}
\label{sec:setup}
\subsection{Experimental Setup} \label{subsec:Setup}

\begin{figure}[t]
    \centering
    \includegraphics[width=\linewidth]{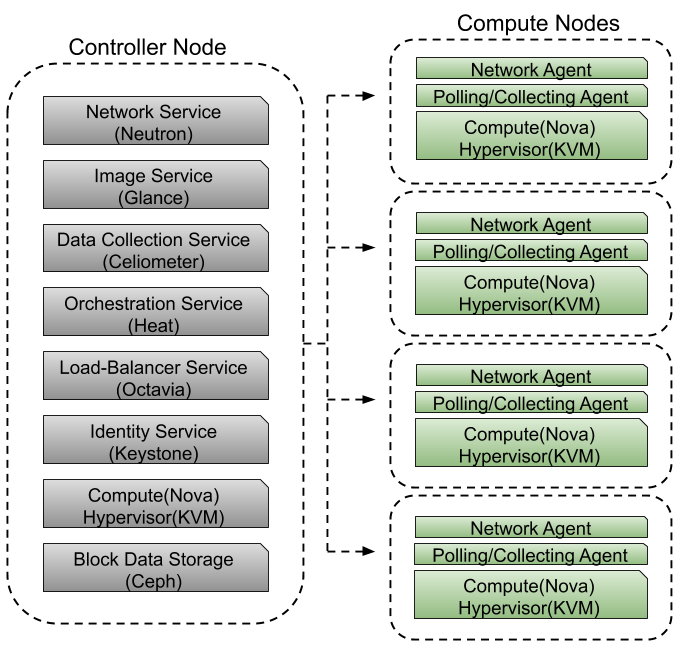}
    \caption{Experimental Cloud Testbed Setup}
    \label{fig:testbed}
\end{figure}

\subsubsection{Testbed} 
Simulating real world data is an imperative aspect to creating reliable malware prediction models. In order to achieve this, we set up a cloud environment with traffic to emulate real world cloud behavior. OpenStack, a popular cloud platform, was installed on the testbed and consisted of a single control node and four compute nodes as shown in Figure \ref{fig:testbed}. The control node is responsible for tasks such as the dashboard, storage, network, identity, and computing. The compute nodes only handle computing services, and each compute node is also supplied with agents for networking, polling, and collecting.
Allowing malware to behave naturally is also key in the data collection process. To ensure the malware was able to behave as \textit{intended}, all of the experiments to collect system features were conducted on machines that were connected to the Internet. This was done since some malware has the capability to detect a closed environment such as a sandbox, a limitation of dynamic malware analysis approaches. If a closed environment is detected, the malware may try to act as a benign process or may cease its behaviour as to not be detected. The Internet connection ensures that malware is able to communicate with its necessary command and control servers needed for malware to act in some cases. In addition, all firewalls and antivirus were disabled.

\begin{figure}[t]
    \centering
    \includegraphics[width=\linewidth]{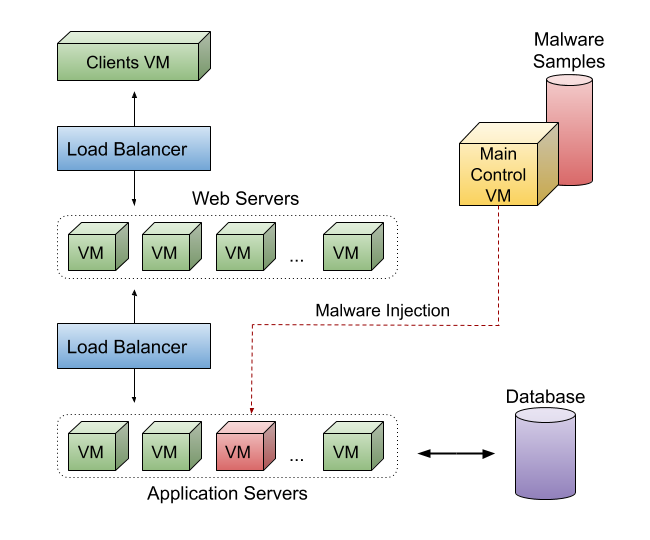}
    \caption{Experiment Deployment}
    \label{fig:experiment}
\end{figure}

\subsubsection{Malware Samples}
The malware that was injected during our experiments were obtained from VirusTotal\footnote{https://www.virustotal.com/gui/}. 113 samples were gathered in total and were chosen at random consisting of diverse malware families such as DoS, Backdoor, Trojan, Virus, among others.

\subsubsection{Experiment Deployment}
In order to simulate a real world scenario, a three-tiered web architecture was used as shown in Figure \ref{fig:experiment}. This architecture consisted of web-servers, application servers, and a database server. A front load balancer is tasked with handling and distributing clients requests to the appropriate web servers.
An internal load balancer is used to connect web servers to the application servers and to distribute requests among the application servers, and the application servers are all connected to a single database server. An auto scaling policy is also utilized which is based on the average CPU usage and is applied independently to both the web and application servers. The policy states that if the average CPU utilization of all VMs belonging to the web or application layer exceeds 70\%, new VMs are created and attached to the corresponding load balancer. Inversely, if the average CPU utilization falls below 40\%, VMs are deleted to reduce resource usage. During our experiments, anywhere between 2 to 10 servers were spawned at each layer, depending on the overall traffic load. In order to uphold integrity, the traffic/requests were generated based on an ON/OFF Pareto distribution, this is done to mimic the realistic dynamic behavior of cloud infrastructures. 
A main control VM is used for keeping the malware executables in the database, injecting a single malware sample into an application server at a specific time, and deploying/destroying the experiment stack. OpenStack Heat orchestration service is used to deploy/destroy experiment stacks using \textit{yaml} scripts.

\begin{figure}[t]
    \centering
    \includegraphics[width=\linewidth]{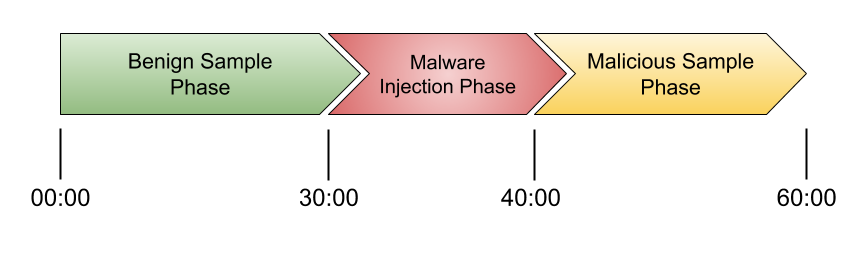}
    \caption{Experiment Timeline}
    \label{fig:timeline}
\end{figure}

\subsubsection{Unique Processes}
We collect system features (e.g. memory, cpu, input/output etc.) from all process that are running on the VM at certain times. Many of these processes are short lived, and also have their process IDs reassigned by the operating system. Due to these characteristics, it is often difficult to analyze the behavior of these processes. To remedy this, we utilize a method known as \textit{unique processes} inspired by \cite{abdelsalam2018malwaremalwarecnn} which reduces dynamism. Traditional operating systems identify processes with a \textit{pid}, whereas unique processes consider the actual behavior of a process and is identified using a tuple of two elements, \textit{process name} and \textit{the command used to run the process}. Processes that share common values in both fields within the tuple are clustered by taking the average of their measurements. By using this approach, we are also able to reduce the total number of processes within a given sample.

\subsubsection{Data Collection}
A total of 113 experiments were conducted each using a different malware executable for a total of 40,680 samples. 
Each of the experiments lasted for 1 hour and is split up into benign and infected phases as shown in Figure \ref{fig:timeline}. The first 30 minutes is the benign phase, during this time there is no malware injected into the machine. Between minute 30 and 40, a single malware is injected into one of the application servers. The malware injection and execution times varies which adds a more dynamic nature to the experiments and ensures that a rigid injection and execution would not skew the results. Minute 40 is referred to as the malicious phase. During this time, malware is openly running on the machine. Data samples are collected every 10 seconds, this results in a total of 360 samples in total for each experiment which are stored in the database. After the experiment is completed, the main control VM destroys the entire experiment stack in order to prevent contamination between experiments.

\subsubsection{Model Training}
We used the scikit-learn\footnote{scikit-learn. https://scikit-learn.org/stable/} library for implementing our classical machine learning models, and Keras\footnote{Keras. https://keras.io/}, a high-level API that runs on top of TensorFlow\footnote{Tensorflow. https://www.tensorflow.org/}, for implementing our CNN model.
All the ML models were trained on a Windows machine equipped with an AMD Ryzen 5 2600 processor and 16 GB of RAM.

\subsection{Evaluation} \label{subsec:Evaluation}
We use five standard metrics to measure the performance of different models, accuracy, precision, recall, and F1 as defined below where TP (True Positive), TN (True Negative), FP (False Positive), and FN (False Negative).
\begin{align*}
    Accuracy &= \frac{TP+TN}{TP+TN+FP+FN}
\end{align*}
\begin{align*}
    Precision &= \frac{TP}{TP+FP}
\end{align*}
\begin{align*}
    Recall &= \frac{TP}{TP+FN}
\end{align*}
\begin{align*}
    F1 ~Score &= 2 \times \frac{Precision \times Recall}{Precision + Recall}
\end{align*}\\

\section{Results} \label{subsec:Results}

\begin{table}[!t]
    \caption{Detailed Performance Results for the Different ML Models}
    \centering
    \begin{tabular}{c | c| c | c | c}
    \hline
    Model & Accuracy & Precision & Recall & F1 \\ [0.5ex] 
    \hline
    
    CNN & 92.9\%&100\%&84.6\%&91.5\% \\
    SVC & 87.56\%&86.2\%&80.91\%&83.47\% \\
    RFC & 89.36\%&99.71\%&72.80\%&84.15\% \\
    KNN & 72.34\%&66.6\%&57.67\%&61.81\% \\
    GBC & 81.47\%&75.22\%&77.87\%&76.57\% \\
    GNB & 58.09\%&48.06\%&98.57\%&64.61\% \\
    
    \hline
    \end{tabular}
    \label{table:results}
\end{table}

Table \ref{table:results} shows the performance metric scores from each of our models. In regards to overall performance, the CNN model outperformed every other model with an F1 score of 91.5\%. We use the F1 score as a basis for overall performance since it takes the precision and recall metrics into consideration and is therefore better at describing a model’s overall performance than the accuracy metric. The RFC and SVC models had the next highest F1 scores with 84.15\% and 83.47\% respectively. The GBC, GNB, and KNN models did not perform as well with F1 scores of 76.57\%, 64.61\%, and 61.81\%. 

The recall metric quantifies the percentage of the infected samples that were detected, and the model that scored the highest in this metric was the GNB model with a score of 98.57\%. However there is a caveat to this, its precision score, which measures how many of the samples that were labeled as infected were actually infected, was very low with a score of 48.06\%. The CNN model achieved a recall score of 84.6\% and a precision score of 100\%. The next best model in regards to recall was the SVC model with a score of 80.91\% and a precision score of 86.2\%. The RFC model achieved a recall score of 72.80\% but also had a very high precision score of 99.71\%. The KNN model performed the worst, which was indicated earlier by the low F1 score, with a recall score of 57.67\% and a precision score of 66.6\%.
The findings when analyzing the accuracy scores are consistent with our other findings, the CNN model outperforms the other models with the RFC and SVC models not too far behind.

\begin{figure}[]
    \centering
    \includegraphics[width=\linewidth]{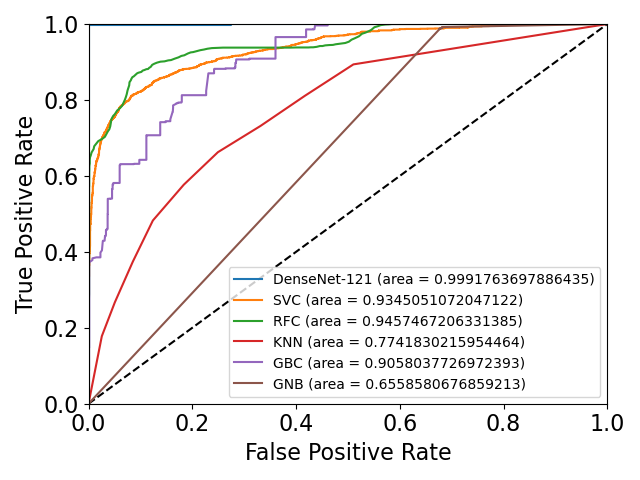}
    \caption{Receiver Operating Characteristic (ROC) Curves}
    \label{fig:roc}
\end{figure}

Figure \ref{fig:roc} shows the ROC curves and AUC scores for each of our models. Once again, the CNN model has the best AUC score of more than 99\%. The RFC and SVC models follow closely scoring a ~94\% and a ~93\% respectively. The GBC model was able to score and AUC of ~90\% whereas the KNN model scored a ~77\% and the GNB model performed the worst with a score of ~65\%. These results are consistent with the results depicted in Table \ref{table:results}.

Another important metric to consider when comparing the performance of various models is the \textbf{training time} of the model, as shown in Table \ref{table:times}. Neural networks require models that use this methodology be trained for a certain number of epochs. If the model is trained for too many epochs \textit{overfitting} could occur, but if the model is not trained for enough epochs, \textit{underfitting} will occur. As such, we stop training if the validation accuracy is not increasing for specific number of epochs and we choose the model with the highest validation accuracy achieved during these epochs. The CNN model took 1683 seconds to train and required a total of 32 epochs. Other models do not train based on a certain number of epochs, therefore, the same technique doesn't apply when training the remainder of the models. In such case, the SVC model took the longest to train by far at 989 seconds. The GBC model took 167 seconds, the KNN model took 28 seconds, the RFC model took 20 seconds, and the GNB model only took 2 second to train. Note that the reported times solely include training time, excluding the time it takes to read and load the data samples.

\section{Comparison and Analysis}
\label{sec:analysis}

\begin{table}[!t]
    \caption{Time Cost for the Models}
    \centering
    \begin{tabular}{c | c | c }
    \hline
    Model & Time to Train (s) & Detection Time (ms) \\ [0.5ex] 
    \hline
    
    CNN & 1683 & 164 \\
    SVC & 989 & 11 \\
    RFC & 20 & 3900 \\
    KNN & 28 & 118 \\
    GBC & 167 & 40 \\
    GNB & 2 & .9 \\
    
    \hline
    \end{tabular}
    \label{table:times}
\end{table}

\begin{figure*}[!t]
    \centering
    \includegraphics[width=0.8\linewidth]{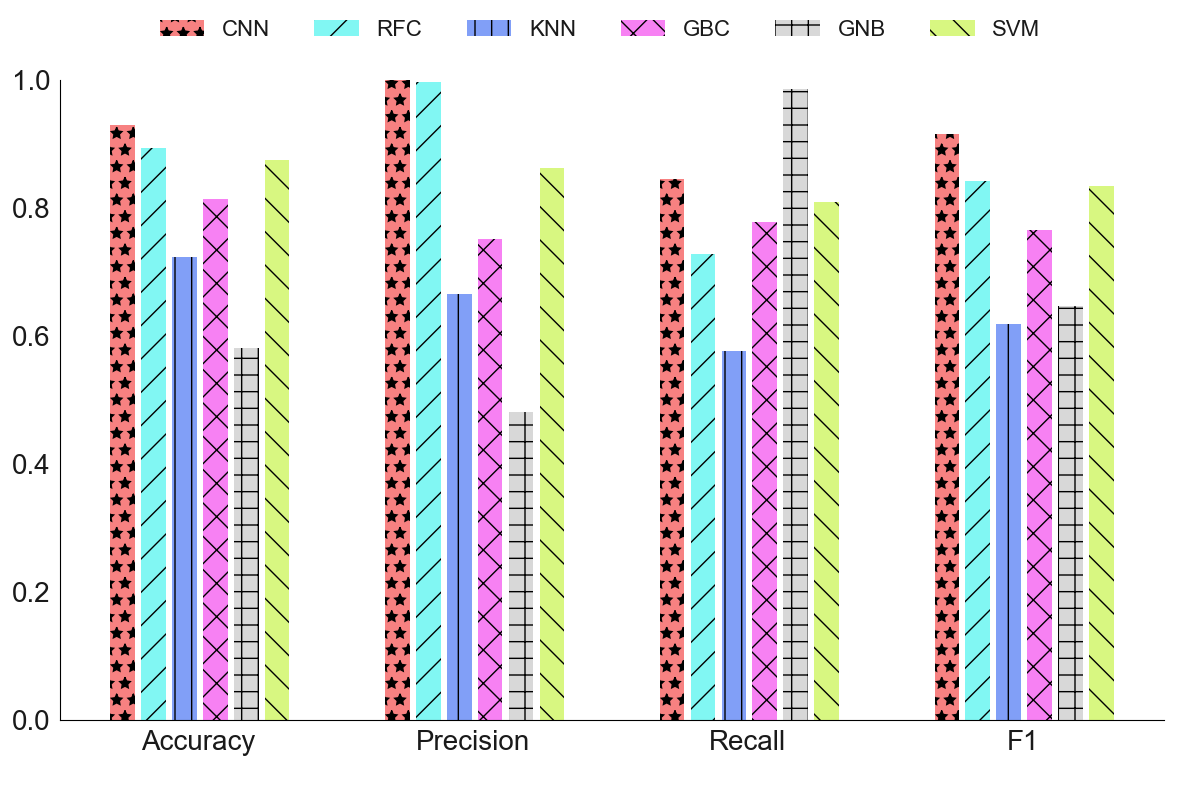}
    \caption{Performance Metrics Comparison for Different Machine Learning Models}
    \label{fig:bar}
\end{figure*}

The superior metrics generated by the CNN model clearly indicate that this model is the best suited for our use case. The comparison of all the models can be found in figure \ref{fig:bar}. The CNN model was able to detect 84.6\% of all of the infected samples while also not falsely labeling benign samples as infected. In malware detection, while detecting every instance of malware is ideal, a high number of false positives can be just as much of a disruption as malware. That is why the high recall rate achieved by the GNB model is not as promising as it may seem. Its extremely low precision score of 48.06\% indicates that this model generated a high number of false positives. In fact, ~52\% of the samples labeled as infected were actually benign. In a business case, having a model that generated these many false positives could hinder day to day activities by labeling essential, benign processes as malware. The SVC and RFC models produced similar accuracy and precision scores but differed in the recall and precision metrics. The SVC model was able to detect ~80\% of the total infected samples, but also has a lower precision score of ~86\%. The RFC model had a lower recall score of ~72\% but generated nearly no false positives with a precision score of ~99\%. These models will need security professionals to make a decision when implementing a malware detection system. There could be a situation where a company is willing to have a lower recall rate in exchange for nearly no false positives to ensure that the malware detection implementation does not severely hinder necessary activities. On the other hand, there may be a security critical use case where a high number of false negatives is very detrimental. In that case, it may be beneficial with a system which is able to detect nearly all malwares even if a large portion of those classifications are false positives.

\subsection{Cost Analysis}

The time it takes to train these models, as shown in Table \ref{table:times}, can also have an affect on which use case each model will be suitable for. There is a clear pattern of the more successful models requiring more time to train. The CNN model took the longest to train by far but also outperformed every other model by a convincing margin. The RFC model gave promising results with an F1 score of ~84\% and also only required 20 seconds to train. The RFC model reigns supreme of the SVC model in regards to time since the SVC model required over 900 seconds to train while still yielding similar results to the RFC model. In a general sense, it is usually worth sacrificing some time in order to train models that are as accurate as possible. That is why deploying a deep learning method such as the CNN model is preferred to the faster trained models such as SVC and RFC. Once these models are trained and deployed for online detection, an important aspect is how long will the models take to decide whether a given sample is benign or malicious. Detection time, also found in Table \ref{table:times}, shows how long it took each model to classify a single sample. This metric indicates how fast each model will be able to process the input data and correctly classify an infected sample as such. The GNB model has the fastest detection time, however this is due to this model having a high false positive rate indicated by its low precision score. The RFC is the slowest model by far, this could be due to the input data having to parse through the various trees that make up the RFC which would cause this model to take longer to generate a prediction. A faster detection time is of course preferred, however if a model generates fast predictions that are incorrect, the benefit of a faster model is overshadowed by its inability to produce accurate results. 

\subsection{Overall Analysis}

Despite its longer training time, the CNN model has proven to be the optimal model in our use case. The CNN model was able to achieve the highest metric scores. 
The CNN model achieved high, balanced values between all of the metrics indicating that this model is able to correctly classify our samples as benign or infected, more so than the other models. The CNN model's longer train time can be attributed to its deep architecture. This model utilized the state of the art DenseNet-121 model, indicating that it contains 121 different layers within the network. 

\section{Limitations}
\label{sec:limitation}

One limitation of our work is due to the relatively small number of malware samples used. We conducted 113 different experiments each with a different type of malware, but conducting more experiments with a wider range of malware could give us a better look into how malware affects the behavior of VMs in a cloud environment. Another limitation lies in the assumption that each VM can only be infected by a single malware, which helps in simplifying our analysis. In practice, a machine can be infected with multiple malware at the same time. That being said, our work aims to provide a fundamental step towards a more rigorous and complex analysis for multiple malware infection. Further work is needed in order to determine if our approach would be feasible given a situation where a VM is infected by multiple malwares. 

In addition, the use of the unique processes approach could allow malware behavior to go unnoticed. Since a unique process sample is the average of a collection of processes sharing the same name and command line, a malware that mimics these same attributes will be counted within the average of this sample. 
This is a common drawback to methodologies that utilize meta-stats (e.g. average, standard deviation, etc.) However, the drawback of using meta-stats is confined to each unique process independently. This makes our approach partially immune to the meta-stats limitation as opposed to other approaches that use meta-stats of the entire system.

\section{Conclusion}
\label{sec:conclusion}
In this paper we analyzed a variety of machine learning methods in order to determine which method is best for online malware detection in cloud. We find that, although it takes the longest to train, the DenseNet-121 (CNN) model has the best overall performance. The SVC and RFC models produced promising results that are not too far behind those produced by the CNN model, as well as being much quicker to train. However, when it comes to malware detection, taking the time to train a more accurate model is mostly preferred. The remaining models, KNN, GBC, and GNB, simply could not compete with the others. Due to the CNN model's success, it can be concluded that deep learning models are more adept at detecting malware within our dataset in cloud IaaS.

\section*{Acknowledgement}
This research is partially supported by NSF Grants 2025682 at Tennessee Technological University and 2025686 at Manhattan College.

\bibliographystyle{./bibliography/IEEEtran}
\bibliography{./bibliography.bib}


\end{document}